\documentclass{article} 
\usepackage{nips13submit_e,times}
\usepackage{hyperref}
\usepackage{url}
\usepackage{bm}
\usepackage[dvipdfmx]{graphicx}
\usepackage{wrapfig}

\title{Sequential Monte Carlo Inference of \\ Mixed Membership Stochastic Blockmodels \\ for Dynamic Social Networks}

\author{
Tomoki Kobayashi, \ Koji Eguchi\\
Graduate School of System Informatics, Kobe University\\
1--1 Rokkodaicho, Nada, Kobe 657--8501, Japan\\
\texttt{kobayashi@cs25.scitec.kobe-u.ac.jp} \\
\texttt{eguchi@port.kobe-u.ac.jp} \\
}

\nipsfinalcopy 

\begin{document}

\maketitle
\begin{abstract}
Many kinds of data can be represented as a network or graph. It is crucial to infer the latent structure underlying such a network and to predict unobserved links in the network. Mixed Membership Stochastic Blockmodel (MMSB) is a promising model for network data. Latent variables and unknown parameters in MMSB have been estimated through Bayesian inference with the entire network; however, it is important to estimate them online for evolving networks. In this paper, we first develop online inference methods for MMSB through sequential Monte Carlo methods, also known as particle filters. We then extend them for time-evolving networks, taking into account the temporal dependency of the network structure. We demonstrate through experiments that the time-dependent particle filter outperformed several baselines in terms of prediction performance in an online condition.
\end{abstract}

\section{Introduction}
Many problems can be represented as networks or graphs, and demands for analyzing such data have increased in recent years. Specifically, it is crucial to infer the latent structure underlying such a network and to predict unobserved links in the network. One promising approach to such problems is latent variable network modeling~\cite{Goldenberg09}. The latent variable network models can be mainly categorized into two kinds. One is hard clustering approaches, such as Stochastic Block Models (SBM)~\cite{Nowicki01} and its variants, which assume each node is assigned to a single cluster or group. On the basis of this assumption, the probability of generating a link from every node in one cluster to another cluster is always the same. In an extension of SBM, Infinite Relational Model (IRM)~\cite{Kemp06} assumes the infinite number of clusters. The other is soft clustering approaches, such as Mixed Membership Stochastic Blockmodels (MMSB)~\cite{Airoldi08}. MMSB assumes that each node is represented as a mixture of multiple latent groups, and that every link is generated in accordance with a Bernoulli distribution associated with each pair of latent groups. MMSB has been successfully applied as social network analysis and protein-protein interaction prediction~\cite{Airoldi08}.

Latent variables and unknown parameters in MMSB have been estimated by variational Bayesian inference or collapsed Gibbs sampling with the {\em entire} network. Those are called batch inference algorithms, requiring significant computational time. However, it is important to estimate them online for evolving networks. In the scenario where nodes or links are sequentially observed over time, it is not realistic to use a batch inference algorithm every time a node or a link is observed, so an online inference algorithm is more appropriate in this case.

For topic models~\cite{Blei03} with text data, various prior studies have explored online inference~\cite{Banerjee07,Canini09,Hoffman10}. 
However, online inference for latent variable network models has not been explored, to the best of our knowledge. In this paper, we propose online inference algorithms for MMSB: incremental Gibbs sampling method and particle filter. Furthermore, we propose another online inference algorithm that dynamically adapts the changes in structure within a network, in the framework of a particle filter. We demonstrate through experiments that these inference methods work effectively for evolving networks. The contributions of this paper are (1) online inference methods for MMSB and (2) novel online inference methods that take into account time dependency in latent structure of evolving networks. 

\begin{figure}[t]
\begin{tabular}{cc}
\begin{minipage}{0.45\hsize}
\begin{center}
\includegraphics[width=\hsize]{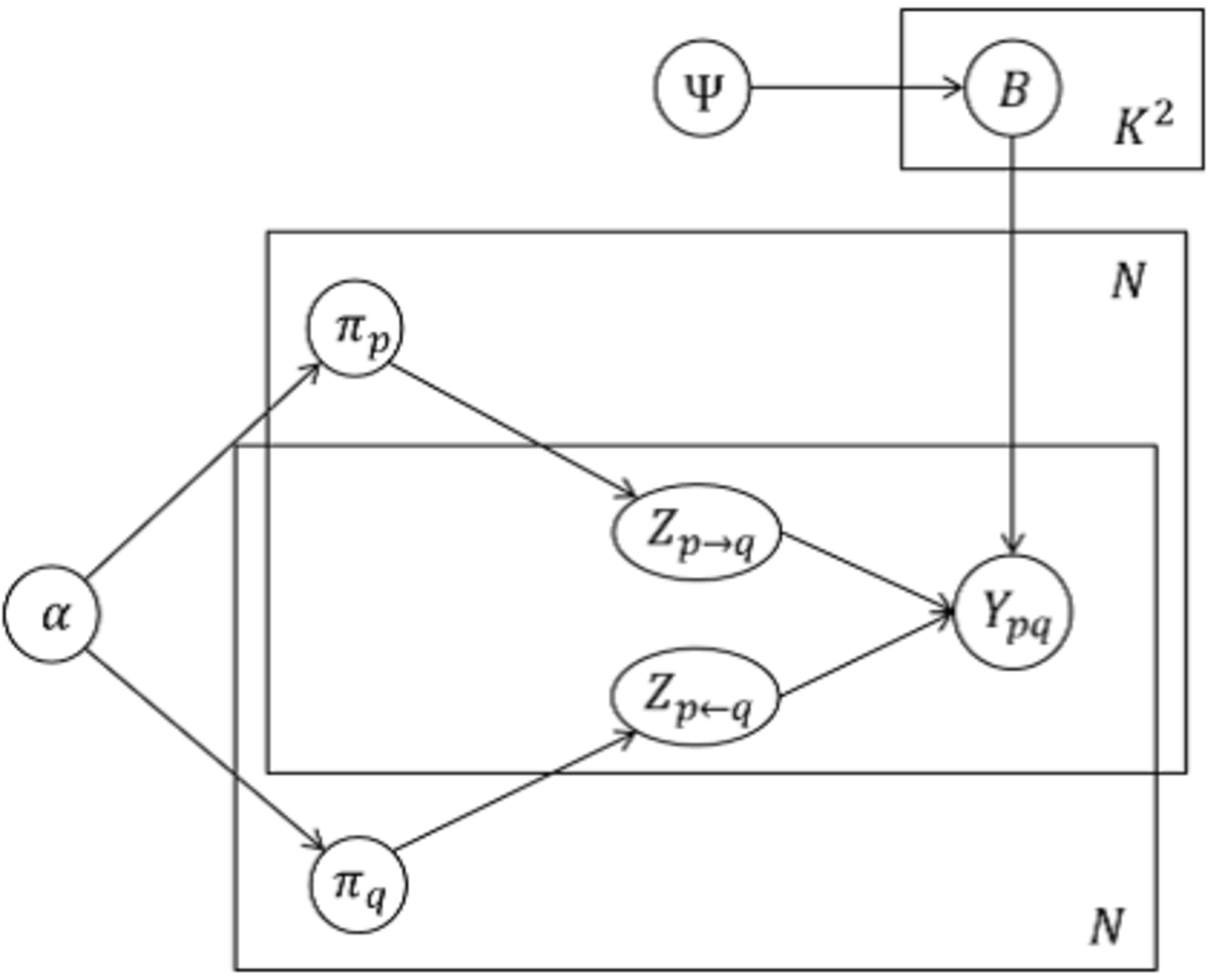}
\caption{Graphical model of MMSB.}
\label{fig:mmsb}
\end{center}
\end{minipage}&
\begin{minipage}{0.45\hsize}
\begin{center}
\small
\scalebox{0.90}{
\begin{tabular}{@{\extracolsep{-\tabcolsep}}rl}\hline
\multicolumn{2}{l}{{\bf Algorithm:} batch Gibbs sampler for $N \times N$}\\
\hline
1:&initialize group assignment randomly for $N \times N$ \\
2:&for iteration=1 to $S$ do\\
3:&\hspace*{1em}for $p$=1 to $N$ do\\
4:&\hspace*{2em}for $q$=1 to $N$ do\\
5:&\hspace*{3em}draw $\mathbf{z}_{p \to q}$ from \\
  &\hspace*{4em}$P({\bf z}_{p \to q}|\mathbf{Y},\mathbf{Z}^{\neg(p,q)}_\to,\mathbf{Z}^{\neg(p,q)}_\gets,{\bm \alpha},{\bm \psi})$\\
6:&\hspace*{3em}draw $\mathbf{z}_{p \gets q}$ from \\
  &\hspace*{4em}$P({\bf z}_{p \gets q}|\mathbf{Y},\mathbf{Z}^{\neg(p,q)}_\to,\mathbf{Z}^{\neg(p,q)}_\gets,{\bm \alpha},{\bm \psi})$\\
7:&\hspace*{2em}end for\\
8:&\hspace*{1em}end for\\
9:&end for\\
10:&complete the posterior estimates of ${\bm \pi}$ and ${\bf B}$\\
\hline
\end{tabular}
}
\end{center}
\caption{Pseudo codes of batch Gibbs sampler.}
\label{fig:batch-code}
\end{minipage}
\end{tabular}
\end{figure}

\section{Mixed Membership Stochastic Blockmodels}
\label{sec:mmsb}
The Mixed Membership Stochastic Blockmodel (MMSB) was proposed by Airoldi et al. This model assumes that each node is represented as a mixture of latent groups, and that every link is generated in accordance with a Bernoulli distribution associated with each pair of latent groups. 

At first, we summarize the definitions used in this paper. 
We represent a simple directed graph as $\mathbf{G}=(\mathbf{N},\mathbf{Y})$, where $(p,q)$ element in an adjacency matrix $\mathbf{Y}$ indicates whether a link (or an arc) is present or absent from node (or vertex) $p$ to node $q$ as $Y(p,q) \in \{0,1\}$. 
Each node is associated with a multinomial distribution over latent groups, $\mathbf{Mult}(\bm{\pi}_p)$. Here, $\pi_{p,g}$ represents the probability that node $p$ belongs to group $g$. Therefore, a single node can be assigned with a different group for each connected link from that node. Supposing that the number of groups is $K$, the relationship between any pair of groups $(g, h)$ is represented as a Bernoulli distribution, $\mathbf{Bern}(\mathbf{B}_{K \times K})$. The element $B(g, h)$ indicates the Bernoulli parameter corresponding to group pair $(g, h)$, representing the probability that a link is present between a node in group $g$ and another node in group $h$. Given a link from node $p$ to $q$, the indicator vector $\mathbf{z}_{p \to q}$ indicates a latent group assigned to node $p$, and $\mathbf{z}_{p \gets q}$ indicates a latent group assigned to node $q$. These latent group indicator vectors are denoted by $\mathbf{Z}_{\to} = \{\mathbf{z}_{p\to q}|p,q\in \mathbf{N}\}$ and $\mathbf{Z}_{\gets} = \{\mathbf{z}_{p\gets q}|p,q\in \mathbf{N}\}$. In accordance with the above definitions, the generative process of MMSB can be described as follows.
\begin{enumerate}%
\item For each node $p$:
\begin{itemize}%
\item Draw a $K$ dimensional vector of multinomial parameters, ${\bm \pi}_{p} \sim {\bf Dir}({\bm \alpha})$
\end{itemize}
\item For each pair of groups $(g, h)$: 
\begin{itemize}%
\item Draw a Bernoulli parameter, $B(g,h) \sim {\bf Beta}({\bm \psi})$
\end{itemize}
\item For each pair of nodes $(p,q)$
\begin{itemize}%
\item Draw an indicator vector for the initiator's group assignment, ${\bf z}_{p\to q} \sim {\bf Mult}({\bm \pi}_p)$
\item Draw an indicator vector for the receiver's group assignment, ${\bf z}_{p\gets q} \sim {\bf Mult}({\bm \pi}_q)$
\item Sample a binary value that represents the presence or absence of a link, $Y(p,q) \sim {\bf Bern}({\bf z}^T_{p \to q}{\bf B}{\bf z}_{p \gets q})$
\end{itemize}
\end{enumerate}
A graphical model representation of MMSB is shown in Fig.~\ref{fig:mmsb}. The full joint distribution of observed data $\mathbf{Y}$ and latent variables $\bm{\pi}_{1:N}$, $\mathbf{Z}_{\to}$, $\mathbf{Z}_{\gets}$, and $\mathbf{B}$ are given as follows: 
\begin{eqnarray}
&&\nonumber P(\mathbf{Y},{\bm \pi}_{1:N},{\bf Z_\to},{\bf Z_\gets},{\bf B}|{\bm \alpha},{\bm \psi})\\
&&=P({\bf B}|{\bm \psi})\prod_{p,q,p \neq q}P(Y(p,q)|{\bf z}_{p\to q},{\bf z}_{p\gets q},{\bf B})P({\bf z}_{p\to q}|{\bm \pi}_p)P({\bf z}_{p\gets q}|{\bm \pi}_q)\prod_p P({\bm \pi}_p|{\bm \alpha})
\end{eqnarray}

Collapsed Gibbs sampling can estimate latent variables and unknown parameters of MMSB. The algorithm is outlined in Fig.~\ref{fig:batch-code}, where it is converged to posterior distributions. Hereinafter, this inference algorithm is referred to as {\em batch Gibbs sampler}. 
This is similar to a collapsed  Gibbs sampler for estimating LDA (Latent Dirichlet allocation) for text data~\cite{Blei03,Griffiths04}. 

\section{Online Inference Algorithms for MMSB}
\label{sec:onlin_algo}
This section describes how to achieve online inference for MMSB, especially by using the incremental Gibbs sampler and particle filter that were originally developed for text data~\cite{Banerjee07,Canini09}. 

\subsection{Incremental Gibbs Sampler}
\label{subsec:inc}
Canini at al.~\cite{Canini09} developed an incremental Gibbs sampler, by modifying batch Gibbs sampler ---also known as collapsed Gibbs sampler~\cite{Griffiths04}---, for estimating an LDA model for text data in an online setting. 
We further modify it for MMSB for network data in an online setting. 

Given a discrete time series of network data, we first apply the batch Gibbs sampler to the first period of the data. Then, every time the presence or absence of a link is observed, we run the following steps:
\begin{enumerate}
\item When a new link with a old node is observed, we sample a pair of groups in accordance with the full conditional probability with already observed data and their group assignments. 
\item When a new link with a new node is observed, we sample a pair of groups for every pair of a new node and an already observed node. 
\item We update the latent groups for the {\em rejuvenation sequence} $\mathcal{R}(p,q)$ ---i.e., $|\mathcal{R}(p,q)|$ of randomly selected node pairs that were already observed at the time when node pair $(p,q)$ is observed---. 
This step is called {\em rejuvenation}, such as in the literature on particle filters~\cite{Doucet01}. 
\end{enumerate}
The larger the $|\mathcal{R}(p,q)|$, the more accurately posterior distributions can be estimated. However, inference time increases quadratically with $|\mathcal{R}(p,q)|$. If we skip the step of rejuvenation or $|\mathcal{R}(p,q)| = 0$, it is similar to the online inference method that Banerjee et al. developed for LDA~\cite{Banerjee07}. 

\begin{figure}[t]
 \begin{center}
\small
\scalebox{0.90}{
\begin{tabular}{
@{\extracolsep{-\tabcolsep}}rl}\hline
\multicolumn{2}{l}{{\bf Algorithm:} particle filter}\\
\hline
1:&initialize weights $\omega^{(k)} = P^{-1}$ for $k=1,\cdots,P$\\
2:&while(add link $p' \to q'$)\\
3:&\hspace*{1em}for $k=1$ to $P$ do \\
4:&\hspace*{2em}$\omega^{(k)} \ \ *\!= \ \ P^{(k)}(Y(p',q')=1|\mathbf{Y},\mathbf{Z}^{\neg(p',q')}_\to,\mathbf{Z}^{\neg(p',q')}_\gets,{\bm \alpha},{\bm \psi}$)\\
5:&\hspace*{2em}draw $z^{(k)}_{p' \to q'}$ from $P^{(k)}({\bf z}_{p' \to q'}|\mathbf{Y},\mathbf{Z}^{\neg(p',q')}_\to,\mathbf{Z}^{\neg(p',q')}_\gets,{\bm \alpha},{\bm \psi}$)\\
6:&\hspace*{2em}draw $z^{(k)}_{p' \gets q'}$ from $P^{(k)}({\bf z}_{p' \gets q'}|\mathbf{Y},\mathbf{Z}^{\neg(p',q')}_\to,\mathbf{Z}^{\neg(p',q')}_\gets,{\bm \alpha},{\bm \psi}$)\\
7:&\hspace*{2em}if($p'$ is new node)\\
8:&\hspace*{3em}for $q=1$ to $N_{current}$ (if $q \neq q'$)\\
9:&\hspace*{4em}draw $z^{(k)}_{p' \to q}$ from $P^{(k)}({\bf z}_{p' \to q}|\mathbf{Y},\mathbf{Z}^{\neg(p',q)}_\to,\mathbf{Z}^{\neg(p',q)}_\gets,{\bm \alpha},{\bm \psi}$)\\
10:&\hspace*{4em}draw $z^{(k)}_{p' \gets q}$ from $P^{(k)}({\bf z}_{p' \gets q}|\mathbf{Y},\mathbf{Z}^{\neg(p',q)}_\to,\mathbf{Z}^{\neg(p',q)}_\gets,{\bm \alpha},{\bm \psi}$)\\
11:&\hspace*{3em}end for\\
12:&\hspace*{2em}end if\\
13:&\hspace*{2em}if($q'$ is new node)\\
14:&\hspace*{3em}for $p=1$ to $N_{current}$ (if $p \neq p'$) do\\
15:&\hspace*{4em}draw $z^{(k)}_{p \to q'}$ from $P^{(k)}({\bf z}_{p \to q'}|\mathbf{Y},\mathbf{Z}^{\neg(p,q')}_\to,\mathbf{Z}^{\neg(p,q')}_\gets,{\bm \alpha},{\bm \psi}$)\\
16:&\hspace*{4em}draw $z^{(k)}_{p \gets q'}$ from $P^{(k)}({\bf z}_{p \gets q'}|\mathbf{Y},\mathbf{Z}^{\neg(p,q')}_\to,\mathbf{Z}^{\neg(p,q')}_\gets,{\bm \alpha},{\bm \psi}$)\\
17:&\hspace*{3em}end for\\
18:&\hspace*{2em}end if\\
19:&\hspace*{1em}end for\\
20:&\hspace*{1em}normalize weights ${\bf \omega}$ to sum to 1\\
21:&\hspace*{1em}if $\| {\bf \omega} \|^{-2} \leq $ESS threshold then\\
22:&\hspace*{2em}resample particles\\
23:&\hspace*{2em}$\omega^{(k)} = P^{-1}$ for $k=1,\cdots,P$\\
24:&\hspace*{1em}end if\\
25:&\hspace*{1em}for $k=1$ to $P$ do \\
26:&\hspace*{2em}for($p'' \to q''$ in $\mathcal{R}(p',q')$) do\\
27:&\hspace*{3em}draw $z^{(k)}_{p'' \to q''}$ from $P^{(k)}({\bf z}_{p'' \to q''}|\mathbf{Y},\mathbf{Z}^{\neg(p'',q'')}_\to,\mathbf{Z}^{\neg(p'',q'')}_\gets,{\bm \alpha},{\bm \psi}$)\\
28:&\hspace*{3em}draw $z^{(k)}_{p'' \gets q''}$ from $P^{(k)}({\bf z}_{p'' \gets q''}|\mathbf{Y},\mathbf{Z}^{\neg(p'',q'')}_\to,\mathbf{Z}^{\neg(p'',q'')}_\gets,{\bm \alpha},{\bm \psi}$)\\
29:&\hspace*{2em}end for\\
30:&\hspace*{1em}end for\\
31:&end while\\
32:&complete the posterior estimates of ${\bm \pi}$ and ${\bf B}$\\
\hline
\end{tabular}
}
\hspace*{5.0\baselineskip}\\
\end{center}
\hspace*{25.0\baselineskip}\\
\caption{Pseudo codes of particle filter.}
\label{fig:pf-code}
\end{figure}

\subsection{Particle Filter}
The particle filter is also known as a sequential Monte Carlo method~\cite{Doucet01}. The inference is achieved by the weighted average of multiple particles, each of which estimates latent group assignments for observed node pairs differently at the same time. Here, the estimation with each particle is performed by following the three steps in an incremental Gibbs sampler, as described in Section~\ref{subsec:inc}. The weight of each particle is assumed to be proportional to the likelihood of generating observed links by the particle. When the variance of the weight is larger than a threshold ---referred to as the effective sample size (ESS) threshold---, resampling is performed to create a new set of particles. We employed a simple resampling scheme that draws particles from the multinomial distribution specified by the normalized weights. After the resampling, the weights are then reset to $P^{-1}$, where $P$ indicates the number of particles. 

The algorithm of the particle filter is outlined in Fig.~\ref{fig:pf-code}.
Note that, in this figure, lines from 5 to 18 correspond to the new link group assignment steps in an incremental Gibbs sampler, and lines from 26 to 29 correspond to the rejuvenation step in an incremental Gibbs sampler. 

The posterior distribution of particle filter, $P_{particle}$ is represented as follows:
\begin{eqnarray}
P_{particle}=\sum_k P^{(k)} \times \omega^{(k)}
\label{eq:likelihood-pf}
\end{eqnarray}
where $P^{(k)}$ indicates the posterior distribution of particle $k$. $\omega^{(k)}$ indicates the weight of particle $k$, which is proportional to the likelihood of generating observed links by the particle. 

\section{Time-dependent Algorithms} 
\label{sec:time-dependent-algorithm}%
We have introduced the inference methods for MMSB, which estimate the latent variables and unknown parameters with already observed data, assuming the network data are sequentially observed over time. However, using all the observed data does not always result in accurate estimation. For instance, your movie preferences, which can be expressed by a bipartite graph, may sometimes change over time. As another example, enterprise email communications expressed as a network may be changed in a structure when serious incidents happen in that company. In such cases, more accurate estimation can be achieved by only considering recent observations and disregarding older observations. On the basis of the idea mentioned above, we propose a time-dependent particle filter for MMSB to capture changes in network structure over time. 


We now describe the method more formally and in more detail. Suppose that $L_t$ represents the likelihood of observations at time $t$. We partially disregard the past observations when the following condition is satisfied. 
\begin{eqnarray}
\lambda_t = \frac{L_t}{L_{t-1}}  < \lambda_0 
\label{eg:histrydeletion}
\end{eqnarray}
where $\lambda_t$ indicates the change rate on the likelihood of observations at time interval $t$, and $\lambda_0$ indicates a threshold parameter. We assume that the pattern of observations is changed when the change rate is small. 
When Eq.~(\ref{eg:histrydeletion}) is satisfied, we further compute $\Lambda_{i,t} = (\lambda_{i+1}, \ldots, \lambda_{t-1}, \lambda_t)$, where each component indicates the change rate of likelihood of observations for each time interval from $i$ to $t$, respectively. 
Here, $i$ is the first time interval that the model considers, so when any time interval was discarded previously, $i=1$. We then sample a time interval to be discarded from a multinomial distribution, whose multinomial parameters are estimated in accordance with $\Lambda_{i,t}$. When time interval $\tau$ ($i \le \tau < t$) is sampled, all the time intervals from $i$ to $\tau$ are then discarded. 

Once the time interval to be disregarded $\tau$ is sampled, we run the following procedure: 
\begin{itemize}
\item When a node is adjacent to any observed node at time interval $\tau$, set the corresponding element of adjacency matrix to be 0. 
\item Randomly assign a latent group to each of these node pairs. 
\item When a node is not adjacent to any observed node at time interval $\tau$, assume the node to be {\em unobserved}. 
\end{itemize}

We can apply this algorithm to the incremental Gibbs sampler and particle filter. When we apply it to the particle filter, we compute $\Lambda_{i,t}$ for each particle. Therefore, each particle can make a different decision on whether or not the past observations should be disregard and which time intervals should be discarded. 

\section{Experiments}
\label{sec:experiment}
In this section, we explore the prediction performance of MMSB estimated via online inference algorithms. For experiments, we use time-series network data. 

\subsection{Settings}
\subsubsection{Data}
\label{sec:data}
For experiments, we use a dataset from the Enron email communication archive~\cite{Klimt04a}. The time period of the dataset is 28 months, from December 1999 to March 2002. 
This dataset is the same as that used by Tang et al.~\cite{Tang08}, where emails in certain folders were removed from each user for the use of email classification research. This dataset was further cleaned so that only the users (i.e., email addresses) who send and receive at least five emails are included. 
We only use the relations between users ---we assume a link from a user to another when a user sends at least one email to another---, disregarding the text content of email messages. 
This dataset contains 2356 nodes.

We divide a set of observations (each link of which is observed to be present or absent for a pair of nodes) evenly into five sets, for the use of five-fold cross validation. 
We further divide each set into a test set and a validation set, and the remaining four sets are used as training set. We briefly summarize the training set, validation set, and test set: 
\begin{description}
\item[Training set]\mbox{}
We use the node pairs included in the training set to estimate the model. The observations (each link of which is observed to be present or absent for a pair of nodes) are sequentially added over time. 
\item[Validation set]\mbox{}
We use the node pairs included in the validation set to determine ESS threshold and $\lambda_0$ threshold for particle filters. 
\item[Test set]\mbox{}
We use the test set to evaluate the inference algorithms. We compute the likelihood of the data within the test set at time interval $t$ using the model estimated with the observations within the training set until time $(t-1)$. 
\end{description}

\subsubsection{Inference methods}
\label{sec:inference}
In the experiments, we compare the following algorithms. 
\begin{description}
\item [Batch Gibbs sampler]\mbox{}
For comparison with online inference, we apply the batch Gibbs sampler by varying the number of Gibbs sweeps (iterations) as $S \in \{50, 100, 150, 200, 250\}$. 

\item [Incremental Gibbs sampler]\mbox{}
In the experiment to compare the batch Gibbs sampler, we set the size of rejuvenation sequence $|\mathcal{R}(p,q)| \in \{0, 1K, 5K, 10K, 20K, 30K, 40K\}$. In the other experiments, we set $|\mathcal{R}(p,q)| \in \{0, 10, 100\}$. 
For all the online algorithms, we carried out the batch Gibbs sampling for the first time interval, setting the number of Gibbs sweeps (iterations) to be $S=100$. 

\item [Particle filter]\mbox{}
We performed a grid search for ESS threshold over $\{4, 8, 12, 16, 20\}$ for each $|\mathcal{R}(p,q)|$ setting. We fix the number of particles to 24 in all the conditions. Using the validation set, we determine ESS threshold for each $|\mathcal{R}(p,q)|$ setting. 


\item [Time-dependent incremental Gibbs sampler]\mbox{}
For comparison, we conducted the experiments with the incremental Gibbs sampler in the same manner as the time-dependent particle filter. 

\item [Time-dependent particle filter]\mbox{}
We experiment with $\lambda_0$ threshold in Eq.~(\ref{eg:histrydeletion}) as 1.0, 1.1, 1.2, 1.3, and 1.4 for each $|\mathcal{R}(p,q)|$ setting. Using the validation set, we determine the optimal $\lambda_0$ for each $|\mathcal{R}(p,q)|$ setting. 

\end{description}

\subsubsection{Evaluation metrics}
Suppose that ${\mathcal T} = \{1,\cdots,T\}$ represents discrete time intervals for a target network. 
We then finally evaluate the prediction performance of each model using the sum of rate of increase in terms of test-set log-likelihood: 
\begin{eqnarray}
\frac{1}{T} \sum_{t=1}^T \frac{X(t)-I_0(t)}{|I_0(t)|} 
\label{eq:evaluation-metric}
\end{eqnarray}
where $I_0(t)$ represents test-set log-likelihood with the baseline: incremental Gibbs sampler at time interval $t$ when $|\mathcal{R}(p,q)|=0$. 
$X(t)$ represents test-set log-likelihood with the target inference method at time interval $t$. 
According to the definition of the prediction performance metric given by Eq.~(\ref{eq:evaluation-metric}), the prediction performance of incremental Gibbs sampler is zero when $|\mathcal{R}(p,q)|=0$. The greater the value of Eq.~(\ref{eq:evaluation-metric}), the more effectively the model works compared with the baseline. 

\subsection{Results}

\begin{figure}[t]
 \begin{center}
\includegraphics[width=6cm]{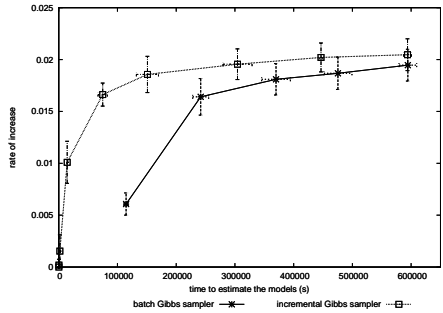}
\caption{Comparison of batch Gibbs sampler and incremental Gibbs sampler. Error bars represent one sample standard deviation.}
\label{fig:result_batch_and_inc}
 \end{center}
\end{figure}

\subsubsection{Batch vs.~online inference}
\label{subsubsec:batch-vs-online}
To understand how online inference methods work, we compare an incremental Gibbs sampler and batch Gibbs sampler, using one of the five sets that were constructed for five-fold cross validation.
We experimented with a machine with a 48-gigabyte memory and a 12-core (24-thread) CPU of 3.06GHz clock speed.
The results are shown in Fig.~\ref{fig:result_batch_and_inc}, where the times (in seconds) to estimate each model and rate of increase of prediction performance in terms of Eq.~(\ref{eq:evaluation-metric}) are demonstrated.
In the case of batch Gibbs sampler, the plots from the left to the right represent $S=50$, $100$, $150$, $200$, and $250$, respectively, indicating that the larger the Gibbs sweeps employed, the longer the required estimation time but the better the prediction performance.
In the case of an incremental Gibbs sampler, the plots from the left to the right represent $|\mathcal{R}(p,q)|=0$, $1K$, $5K$, $10K$, $20K$, $30K$, and $40K$, respectively, indicating that the larger the assumed rejuvenation sequence, the longer the required estimation time but the better the prediction performance. 

As you can see in Fig.~\ref{fig:result_batch_and_inc}, online inference with an incremental Gibbs sampler is much faster than that with a batch Gibbs sampler. 
Both inference methods are almost converged at around 0.018 of prediction performance.
To achieve that prediction performance, the batch Gibbs sampler (when $S=200$) took 475,374 seconds on average, while the incremental Gibbs sampler (when $|\mathcal{R}(p,q)|=10K$) took 151,027 seconds on average: 32\% of that of the batch Gibbs sampler.
Another online inference method, the particle filter, behaves similarly to (though slightly differently from) the incremental Gibbs sampler but very differently from the batch Gibbs sampler. 

Moreover, the batch Gibbs sampler is invoked at every time interval, so it takes more time when we assume finer time intervals. On the other hand, online inference methods such as the incremental Gibbs sampler and particle filter take constant time since we update the latent variables and unknown parameters sequentially, no matter how we define time intervals. For these reasons, online inference is more appropriate for when the target network is sequentially observed. 

\begin{figure}[t]
 \begin{minipage}{0.49\hsize}
\includegraphics[height=4.5cm]{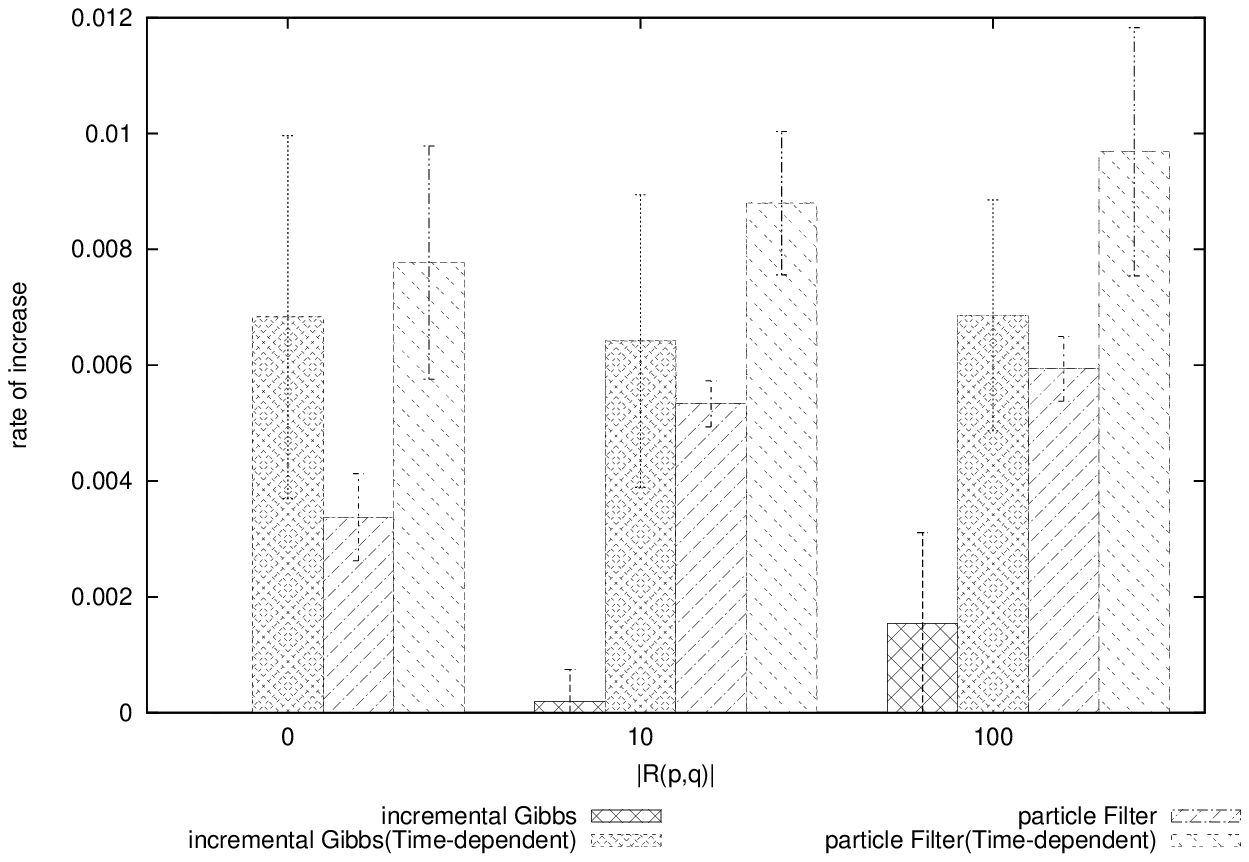}
\caption{Prediction performance of time-dependent inference methods. Error bars represent one sample standard deviation.}
\label{fig:resultA_bP}
 \end{minipage}
 \begin{minipage}{0.06\hsize}

 \end{minipage}
 \begin{minipage}{0.49\hsize}
\includegraphics[height=4.5cm]{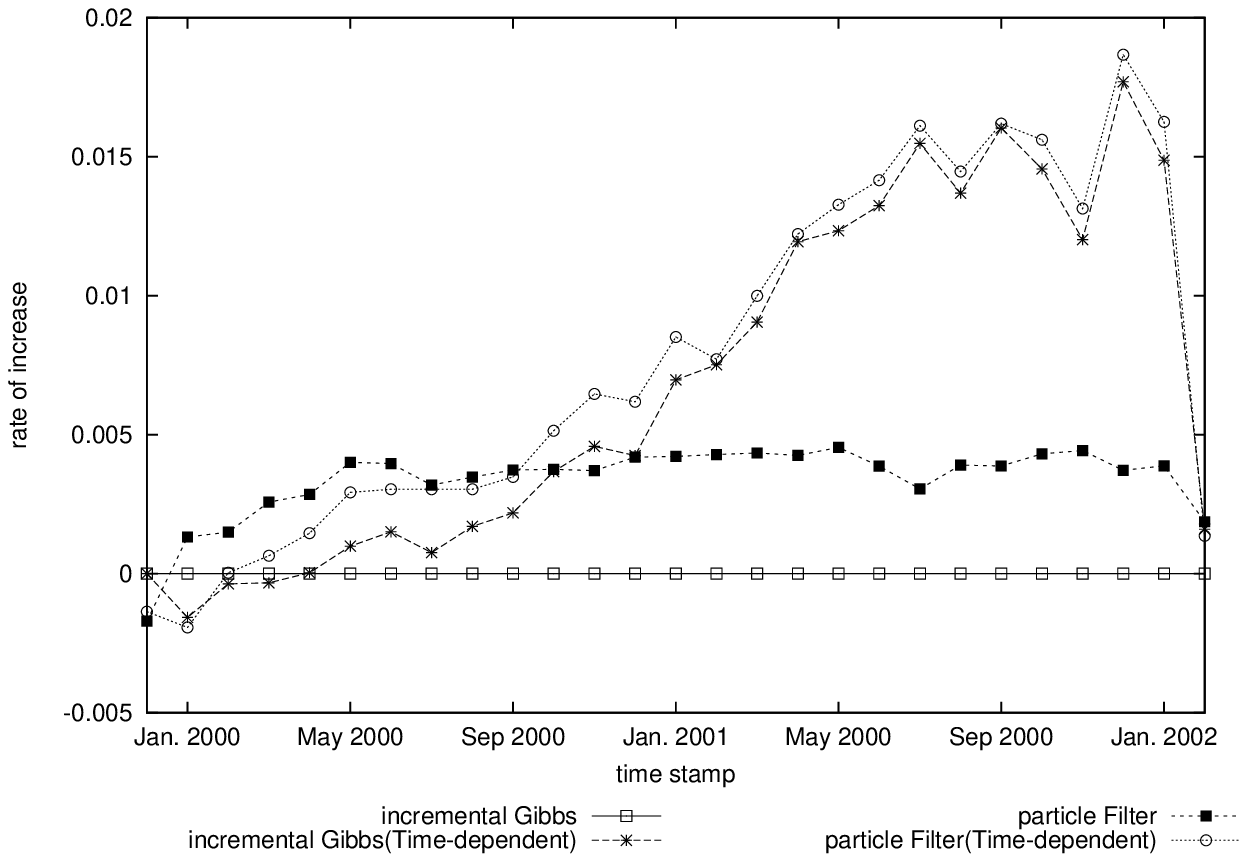}
\caption{Prediction performance of time-dependent inference methods in time series plots.}
\label{fig:resultA_bP_stream}
 \end{minipage}
\end{figure}
\subsubsection{Incremental Gibbs sampler vs.~particle filter}
\label{subsubsec:inc-vs-pf}
Fig.~\ref{fig:resultA_bP} shows the prediction performance of the incremental Gibbs sampler and particle filter using time-independent and time-dependent algorithms, respectively. 
The run time of the particle filter is six times longer than that of incremental Gibbs for the same $|\mathcal{R}(p,q)|$ setting; however, it is much less than that of the batch Gibbs sampler.
The greater the $|\mathcal{R}(p,q)|$, the better the prediction performance but the longer the required time.
As shown in Fig.~\ref{fig:resultA_bP}, the prediction performance of particle filter is greater than that of incremental Gibbs in any $|\mathcal{R}(p,q)|$ setting. 

\subsubsection{Time-dependent algorithms for incremental Gibbs sampler and particle filter}
\label{subsubsec:time-dependent}
As shown in Fig.~\ref{fig:resultA_bP}, the time-dependent algorithm works effectively in both the incremental Gibbs sampler and particle filter. 
Moreover, the prediction performance of the time-dependent particle filter is greater than that of the time-dependent incremental Gibbs sampler.

The Enron Corporation ---an U.S.~energy, commodities, and services company--- experienced a drop in stock market price from around January 2001 and then declared bankruptcy on December 2, 2001. The datasets we used for experiments are based on email communications within that company from December 1999 to March 2002, as we mentioned previously. Therefore, the network structure of email communications should be continuously changed, especially from January 2001 to December 2001. 
We demonstrate whether and how effectively the time-dependent inference methods capture the change in Fig.~\ref{fig:resultA_bP_stream}, by plotting in time order the improvements in prediction performance, $\frac{X(t)-I_0(t)}{|I_0(t)|}$ in Eq.(\ref{eq:evaluation-metric}), on the basis of that of the incremental Gibbs sampler when $|\mathcal{R}(p,q)|=0$. 
As you can see in this figure, the time-dependent particle filter and time-dependent incremental Gibbs sampler outperform the time-independent methods, especially from January 2001. This indicates that the time-dependent inference methods adequately capture the change in structure of the email communications during the crisis at the company, by removing past observations in online inference. 

\section{Conclusions}
\label{sec:conclusion}
In this paper, we proposed online inference methods for Mixed Membership Stochastic Blockmodels (MMSB) that have never been explored. 
Furthermore, we also proposed time-dependent algorithms for the online inference of MMSB, reflecting the change in structure of the network data over time by selectively discarding the past observations when the change occurs. 
We experimented with an email communication dataset to evaluate both the prediction performance and the time required for the estimation. We demonstrated that particle filter improved prediction performance compared with the baselines of the batch Gibbs sampler and incremental Gibbs sampler. 
We also demonstrated that the time-dependent particle filter works more effectively than either the naive particle filter or time-dependent incremental Gibbs sampler. 
More detailed evaluation is left for our future work. 
Applying our inference methods to a nonparametric relational model~\cite{Miller09} is one of the possible directions for the future work.

\noindent {\bf Acknowledgments:}
This work was supported in part by the Grant-in-Aid for Scientific Research (\#23300039) from JSPS, Japan.


\end{document}